\documentstyle[12pt,aps,prc,epsfig,preprint]{revtex}
\tightenlines
\begin{document}

\begin{titlepage}
\title{\vspace*{5mm}\bf
\Large Medium Effects of Low Energy Pions}
\vspace{4pt}

\author{E.~Friedman  \\
{\it Racah Institute of Physics, The Hebrew University,
Jerusalem 91904, Israel\\}}
\vspace{4pt}
\maketitle

\begin{abstract}

Fits of pion-nucleus potentials to large sets of pionic atom data
reveal departures of parameter values from the corresponding free
$\pi N$ parameters. These medium effects can be quantitatively
reproduced by a chiral-motivated model where the pion decay constant
is modified in the medium or by including the empirical on-shell
energy dependence of the  amplitudes. No consistency is obtained
between pionic atoms and the free $\pi N$ interaction when
an extreme off-shell chiral model is used. The role of the size of data sets
is briefly discussed.

$PACS$: 12.39.Fe; 13.75.Gx; 21.30.Fe; 36.10.Gv
\newline
\newline \newline
Corresponding author: E. Friedman, \newline
Tel: +972 2 658 4667, FAX: +972 2 658 6347, \newline
E mail: elifried@vms.huji.ac.il

\end{abstract}

\centerline{\today}
\end{titlepage}

\section{Introduction}
\label{sec:int}

Renewed interest in pionic atoms in general, and in the $s$-wave
part of the pion-nucleus potential in particular, stems from three recent
developments. The first is the experimental observation of `deeply
bound' pionic atom  states in the (d,$^3$He) reaction
\cite{YHI96,GGK00,GGG02,SFG02}, the existence of which was
predicted a decade earlier
\cite{FSo85,TYa88,THY89}. The second is the very accurate measurements
of the shift and width of the 1$s$ level in pionic
hydrogen \cite{SBG01} and in pionic deuterium \cite{CEJ97,HKS98}
which leads to precise values for the $s$-wave
$\pi N$ scattering lengths (see also Ref. \cite{ELT02}).
The third development is the attempt to explain the `anomalous'
$s$-wave repulsion
\cite{BFG97} in terms of a density dependence of the pion decay constant
\cite{Wei01}, or very recently by
constructing the $\pi N$ amplitude near threshold within a systematic
chiral perturbation expansion \cite{KWe01}
and in particular imposing on it gauge invariance \cite{KKW03,KKW03a}.

The so-called anomalous repulsion of the $s$-wave pionic atom potential
is the empirical finding, from fits of optical potentials to pionic atom
data, that the strength of the repulsive $s$-wave
potential inside nuclei is typically twice as large as is expected on
the basis of the free $\pi N$ interaction. This enhancement is, to a 
large extent, due to the value of the potential parameter $b_1$, 
the in-medium
{\it isovector} $s$-wave $\pi N$ amplitude. 
 The possible modification in the medium of this amplitude
 is the topic of the present work. One may define medium
effects in this context as unexpected results obtained from analyses
of experimental data, usually with rather simple models, based in the
present case on the free pion-nucleon interaction.

Section \ref{sect:potl} presents briefly the
standard pion-nucleus potential \cite{EEr66} where
the various parameters are defined. 
Section \ref{sect:con} shows results of fits of the standard potential
to a large-scale (`global') set of data where medium effects are presented
in a quantitative way. Section \ref{sect:W} shows similar results for the
model due to Weise \cite{Wei01} where the pion decay constant is assumed to 
be modified in the medium. An alternative approach where the medium
effects are essentially due to the energy dependence of the amplitudes
is discussed in section \ref{sect:E} both for  a chiral (off shell)
approach and for an empirical (on-shell) approach. 
In section \ref{sect:stat} the statistical significance of the results 
and their relation to the extent of the data base are  discussed
together with consequences of constraining parameter values.
Section \ref{sect:sum} is a summary.

\section{The pion-nucleus potential} \label{sect:potl}

The optical potential $V_{\rm opt}$ used in the Klein-Gordon
equation is of
the standard form due to  Ericson and Ericson \cite{EEr66},

\begin{equation} \label{equ:EE1}
2\mu V_{{\rm opt}}(r) = q(r) + \vec \nabla \cdot \alpha(r) \vec \nabla
\end{equation}
with
\begin{eqnarray} \label{equ:EE1s}
q(r) & = & -4\pi(1+\frac{\mu}{M})\{{\bar b_0}(r)[\rho_n(r)+\rho_p(r)]
  +b_1[\rho_n(r)-\rho_p(r)] \} \nonumber \\
 & &  -4\pi(1+\frac{\mu}{2M})4B_0\rho_n(r) \rho_p(r) ,
\end{eqnarray}

\begin{equation} \label{equ:LL2}
\alpha (r) = \frac{\alpha _1(r)}{1+\frac{1}{3} \xi \alpha _1(r)}
 + \alpha _2(r) ,
\end{equation}
\noindent
where

\begin{equation} \label{equ:alp1}
\alpha _1(r) = 4\pi (1+\frac{\mu}{M})^{-1} \{c_0[\rho _n(r)
  +\rho _p(r)] +  c_1[\rho _n(r)-\rho _p(r)] \} ,
\end{equation}

\begin{equation} \label{equ:alp2}
\alpha _2(r) = 4\pi (1+\frac{\mu}{2M})^{-1} 4C_0\rho _n(r) \rho _p(r).
\end{equation}

\noindent
In these expressions $\rho_n$ and $\rho_p$ are the neutron and proton density
distributions normalized to the number of neutrons $N$ and number
of protons $Z$, respectively, $\mu$ is the pion-nucleus reduced mass
and $M$ is the mass of the nucleon,
 $q(r)$ is referred to
as the $s$-wave potential term and $\alpha(r)$ is referred to
as the $p$-wave potential term. The latter will not be discussed here further
as there are essentially no non-trivial medium effects with this term.
The function ${\bar b_0}(r)$ in Eq. (\ref{equ:EE1s})
is given in terms of the {\it local} Fermi
momentum $k_{\rm F}(r)$ corresponding to the isoscalar nucleon
density distribution:

\begin{equation} \label{equ:b0b}
{\bar b_0}(r) = b_0 - \frac{3}{2\pi}(b_0^2+2b_1^2)k_{\rm F}(r),
\end{equation}
where $b_0$ and $b_1$ are minus the pion-nucleon isoscalar
and isovector effective scattering lengths, respectively.
The quadratic terms in $b_0$ and $b_1$ represent double-scattering
modifications of $b_0$. In particular, the $b_1^2$ term represents
a sizable correction to the nearly vanishing linear $b_0$ term.

The nuclear density
distributions $\rho_p$ and $\rho_n$ are an essential part of the
pion-nucleus potential. Density distributions for protons can  be obtained by
unfolding the finite size of the charge of the proton from charge
distributions obtained from experiments with electrons or muons. This leads to
reliable proton densities,
particularly in the surface region, which is the relevant region for
producing strong-interaction effects in pionic atoms. The neutron distributions
are, however, generally not known to sufficient accuracy and
 we have therefore adopted a
semi-phenomenological approach that covers a broad range of possible
neutron density distributions.
The feature of  neutron density distributions which is most effective
in  determining parameter values of the potential
 is the radial extent, as represented, for
example, by $r_n$, the rms radius of the neutron density. Other features
such as  the detailed
shape of the distribution have only minor effect on values of 
potential parameters, although they do affect the quality of fits. 
For that reason we chose
the rms radius as the prime parameter in the present study, and focus attention
on values of $r_n-r_p$, the difference between the rms radii.
A simple parameterization  adopted here 
that is easy to relate to the results of
relativistic mean field (RMF) calculations \cite{LRR99,FGa03} is

\begin{equation} \label{equ:RMF}
r_n-r_p = \alpha \frac{N-Z}{A} + \eta .
\end{equation}

\noindent In order to allow for possible differences in the shape 
of the neutron
distribution, the `skin' and the `halo' forms of Ref. \cite{FGa03} have
been used. Note that the results of RMF calculations are described very well
\cite{FGa03} by using $\alpha$=1.5 fm and $\eta $= $-$0.03 fm.
This value of $\alpha $ is probably an upper
limit as other sources of information suggest values close to 1-1.2 fm.

\section{Results for  the standard potential} \label{sect:con}

Results obtained from the use of the standard potential 
of the previous section, here marked as `conventional', with 100
data points along the periodic table, from $^{20}$Ne to $^{238}$U,
are shown in Fig.~\ref{fig:con}. 
These data include also the deeply bound
1$s$ states in $^{205}$Pb \cite{GGG02} and in $^{115,119,123}$Sn \cite{SFG02}.
 The figure shows that for $\alpha$=1.5 and for 
the `skin' type of density a very good fit to the data is obtained ($\chi ^2$
per point of 1.7) but the resulting  
value of $b_1$ is more repulsive than the free $\pi N$
value by at least two error bars, or three error bars if the more reasonable
value of $\alpha$=1.25 is adopted as representing the average behaviour of
rms radii of neutron distributions. This enhanced repulsion is a clear
indication of medium effects. Other medium effects in the potential
are likely to be present in the phenomenological two-nucleon absorption 
terms $B_0$ and $C_0$ 
which also have real dispersive parts. These are more 
difficult to handle and we will not discuss these here except for
 noting that the real parts of $B_0$ and $C_0$ are expected not
to exceed in absolute values the corresponding imaginary parts, respectively. 
The anomalous $s$-wave repulsion observed with the conventional potential
is due to the  combined action of the too repulsive
$b_1$ (which enters also quadratically, see Eq. (\ref{equ:b0b}))
and of Re$B_0$ which is also found to be too repulsive compared to the above
expectations (see also section \ref{sect:stat} below).

\section{Medium-modified pion decay constant} \label{sect:W}

The in-medium $s$-wave interaction of pions have been discussed 
recently by Weise \cite{Wei01}
in terms of partial restoration of chiral symmetry in dense matter.
Since $b_{1}$ in free-space is well approximated in lowest
chiral-expansion order by the Tomozawa-Weinberg expression
\cite{Tom66}

\begin{equation}
\label{equ:b1}
b_{1}=-\frac{\mu_{\pi N}}{8 \pi f^{2}_{\pi}}=-0.08~m^{-1}_{\pi} \,,
\end{equation}
then it can be argued that $b_{1}$ will be modified in pionic atoms 
if the pion decay constant
$f_\pi$ is modified in the medium. The square of this decay constant
is given, in leading order,
 as a linear function of the nuclear density,

\begin{equation} \label{eq:fpi2}
f_\pi ^{*2} = f_\pi ^2 - \frac{\sigma }{m_\pi ^2} \rho
\end{equation}
with $\sigma$ the pion-nucleon sigma term.
This leads to a density-dependent isovector amplitude such that $b_1$ becomes

\begin{equation}\label{eq:ddb1}
b_1(\rho) = \frac{b_1(0)}{1-2.3\rho}
\end{equation}
for $\sigma $=50 MeV \cite{San02} and
with $\rho$ in units of fm$^{-3}$.
This model was found \cite{Fri02,Fri02a} to be very successful when tested
against large data sets.

Figure \ref{fig:W} shows results similar to those of Fig. \ref{fig:con}
but with $b_1$ as given by Eq. (\ref{eq:ddb1}). Only the results for the `skin'
shape of the neutron densities are plotted as the `halo' shape always 
produces poorer fits to the data. It is seen from the figure 
that the quality of fits to the data is the same as for the conventional
model, but the values of $b_1$ have shifted now such that for 
$\alpha \sim $ 1.25 they agree with the free $\pi N$ value, thus indicating
that the density dependence of the decay constant could account for the
medium modification of $b_1$.
Using the present model the parameter Re$B_0$ is found to be considerably
less repulsive than in the conventional model, and  within errors
in agreement with expectations \cite{Fri02}.

\section{Energy-dependent amplitudes} \label{sect:E}
Recently, Kolomeitsev, Kaiser and Weise \cite{KKW03} have suggested
that pionic atom data could be reproduced using a pion optical
potential underlain by chirally expanded $\pi N$ amplitudes,
retaining the energy dependence of the amplitudes $b_{0}(E)$ and
$b_{1}(E)$ for zero-momentum ({\bf q}=0) pions in nuclear matter
in order to impose the minimal substitution requirement
$E \to E - V_{c}$, where $V_c$ is the Coulomb potential.
This has the advantage of enabling one to use
a systematic chiral expansion as an input \cite{KWe01}, rather than
singling out the leading order term Eq. (\ref{equ:b1}) for $b_{1}$.

The chiral expansion of the $\pi N$ amplitudes for {\bf q} = 0 at the
two-loop level is well approximated by the following expressions
\cite{KWe01,KKW03}:
\begin{equation}
4\pi \left( {1+{{m_\pi } \over M}} \right)b_{0}(E)\approx
 \left( {{{\sigma -\beta E^2} \over {f_\pi ^2}}+{{3g_A^2m_\pi ^3} \over
{16\pi f_\pi ^4}}} \right)\   ,
        \label{eqn:b0ch}
\end{equation}

\begin{equation}
4\pi \left( {1+{{m_\pi } \over M}} \right)b_{1}(E)\approx
-{E \over {2f_\pi ^2}} \left( {1+{{\gamma E^2} \over {\left( {2\pi f_\pi }
\right)^2}}} \right)\   ,
        \label{eqn:b1ch}
\end{equation}
 $g_{A}$ is the nucleon axial-vector coupling constant,
$g_{A}=1.27$, $\beta$ and $\gamma$ are tuned to reproduce the
threshold values $b_{0}(m_{\pi}) \approx 0$ and $b_{1}(m_{\pi}) =
-0.0885^{+0.0010}_{-0.0021}~~m^{-1}_{\pi}$ \cite{SBG01} respectively.
For $b_{0}$, in view of the accidental cancellations that lead to
its near vanishing we limit our discussion to the $f_{\pi}^{-2}$
term in Eq. (\ref{eqn:b0ch}), therefore choosing
$\beta = \sigma m_{\pi}^{-2}$.

Implementing the minimal substitution requirement in
the calculation of pionic atom observables, the constant parameters
$b_{0,1}$ of the conventionally energy independent optical potential
have been replaced in our calculation \cite{FGa03a} by

\begin{equation}\label{equ:b01}
b_{0,1}(r)=b_{0,1} - \delta _{0,1} ({\rm Re}B + V_c(r)) \,,
\end{equation}
where $\delta _{0,1} = \partial b_{0,1}(E)/\partial E$ is the
appropriate slope parameter at threshold, Re$B$ is the (real)
binding energy of the corresponding pionic atom state and $V_c(r)$
is the Coulomb potential. The constant fit parameters $b_{0,1}$ are
then expected to agree with the corresponding free $\pi N$ threshold
amplitudes if the energy dependence is indeed responsible for the
renormalized values found in conventional analyses.

In addition to the above `chiral' energy dependence for {\it off-shell}
{\bf q} = 0 pions we also present results for the empirically known
{\it on-shell} $\pi N$ amplitudes, when the pion energy $E$ and its
three-momentum {\bf q} are related by $E^2 = m_{\pi}^2 + {\bf q}^2$.
This choice corresponds to the original suggestion by Ericson and
Tauscher \cite{ETa82} to consider the effect of energy dependence
in pionic atoms. Ericson subsequently \cite{Eri94} pointed out that,
for strongly repulsive short-range $NN$ correlations, the on-shell
requirement follows naturally from the Agassi-Gal theorem \cite{AGa73}
for scattering off non-overlapping nucleons. The corresponding
$\pi N$ amplitudes are denoted  as `empirical' and are taken from the
SAID data base \cite{SAID}.

Figure \ref{fig:E} shows results for the two kinds of energy-dependence
mentioned above. The left panels show that implementing the chiral
off-shell energy dependence leads to very significant deterioration
in the fit to the data both for pionic atoms (lower part) and for
the free $\pi N$ interaction (upper part). Note also that the
minimum in the $\chi ^2$ curve occurs at unacceptably large value
of the parameter $\alpha$. In contrast, the right hand side of the figure
shows good agreement both for pionic atoms and for the free $\pi N$ 
 interaction, in accordance with the original suggestion made by  
Ericson and Tauscher \cite{ETa82}.
Perhaps it is not too surprising that the off-shell {\bf q}=0 
nuclear matter approach
is inadequate when the pion-nuclear optical potential that generates 
pionic-atom wavefunctions is being considered.

\section{Statistical considerations and data sets} \label{sect:stat}

The results discussed so far have all been based on the `global 3' data
set of Ref. \cite{FGa03} which consists of 100 data points. Any
conclusion regarding medium effects  must obviously
be linked to the uncertainties in the extracted parameter values, 
and these
invariably depend on the size of the data sets used. In addition the
uncertainties may depend critically on assumptions made in the analysis
such as assigning fixed values to some parameters. These points are
demonstrated in Table \ref{tab:stat}.

The different columns of the table refer to various data sets, ranging
from 120 points for nuclei between $^{12}$C and $^{238}$U to just 20
points where the deeply bound states provide half of the data. The top
half of the Table is for unrestricted $\chi ^2$ fits whereas the lower
part is for corresponding fits  with the parameter Re$B_0$ set to zero.
All the results in this table are for a conventional potential, i.e.
without explicit medium effects such as density-dependence or 
energy-dependence of the amplitudes, and for neutron densities calculated
for $\alpha $=1.5 (Eq. ({\ref{equ:RMF})). From the top half of the table
it is seen that the values of $b_0$ are essentially in agreement with
the free $\pi N$ value, within their rather large uncertainties and that 
the values of $b_1$ disagree with the corresponding free $\pi N$ value
only for the two large data sets since the uncertainties become too
large for the two smaller data sets to reach such a conclusion. 
This distinction becomes even sharper
when the uncertainty due to the neutron distributions is also included
\cite{FGa03}. Another feature of the top half of the table is that
the parameter Re$B_0$ is distinctly different from zero, although its
uncertainty is not small, only when the large data bases are being used.
For the two smaller data sets its value is consistent with zero.

Some recent papers insist on using the value of Re$B_0$=0 
\cite {GGK00,GGG02,SFG02} and the
lower part of the table addresses the consequences of such a constraint.
The first consequence is that the uncertainty
of the parameter $b_0$ becomes much smaller than before and its values,
in all cases, deviate sharply from the free $\pi N$ value. This conclusion
remains valid also for all the types of medium effects for $b_1$ considered
in the present work. 
The second consequence is that for the large data sets the constraint
Re$B_0$=0 indeed leads to a significant deterioration in the quality
of  fits, as evidenced by the increase of $\chi ^2$. Recall that
the natural unit for such increases is the value of $\chi ^2/ F$,
the $\chi ^2$ per degree of freedom. For the large data sets the 
increase is 10 such units, which is very significant. 
Note also that  the values of $b_1$ do {\it not} depend on 
whether or not Re$B_0$ is set to zero. Similar
results are obtained for the other models for $b_1$ discussed above,
but with values of $b_1$ in agreement with what is displayed in the 
 figures.
It is concluded that analyses of reduced data sets with the constraint
of Re$B_0$=0 must lead to unreliable results. In fact, analysing 
{\it only} the deeply bound states for $^{115,119,123}$Sn and
$^{205}$Pb, we obtain excellent fits to the data with values of 
$b_1$ anywhere between $-$0.07 and $-$0.13 $m_\pi ^{-1}$.

\section{Summary} \label{sect:sum}
We have shown that global fits to large sets of data on strong interaction
effects in pionic atoms in terms of a theoretically-motivated optical
potential lead to very good description of the data ($\chi ^2 /N \sim $1.7)
with well-determined $s$-wave isovector amplitude that differs from
the corresponding free $\pi N$ amplitude by  three standard deviations.
This difference indicates modification of the interaction in the medium
and it was shown that a chiral-motivated model where the pion decay constant
is modified in the medium is capable of reproducing the medium effects.
An alternative approach where the empirical on-shell energy dependence
of the $\pi N$ amplitude is included within the minimal substitution
$E \to E - V_{c}$ was also shown to be successful. In contrast, the
fully off-shell chiral energy dependence of the amplitudes fails badly.
The importance of unrestricted and unbiased fits to large scale data sets
has been demonstrated.

\vspace{9mm}

I wish to thank A. Gal for many fruitful discussions throughout the whole 
project. This research was partially supported by the
Israel Science Foundation grant No. 131/01.

\begin{figure}
\epsfig{file=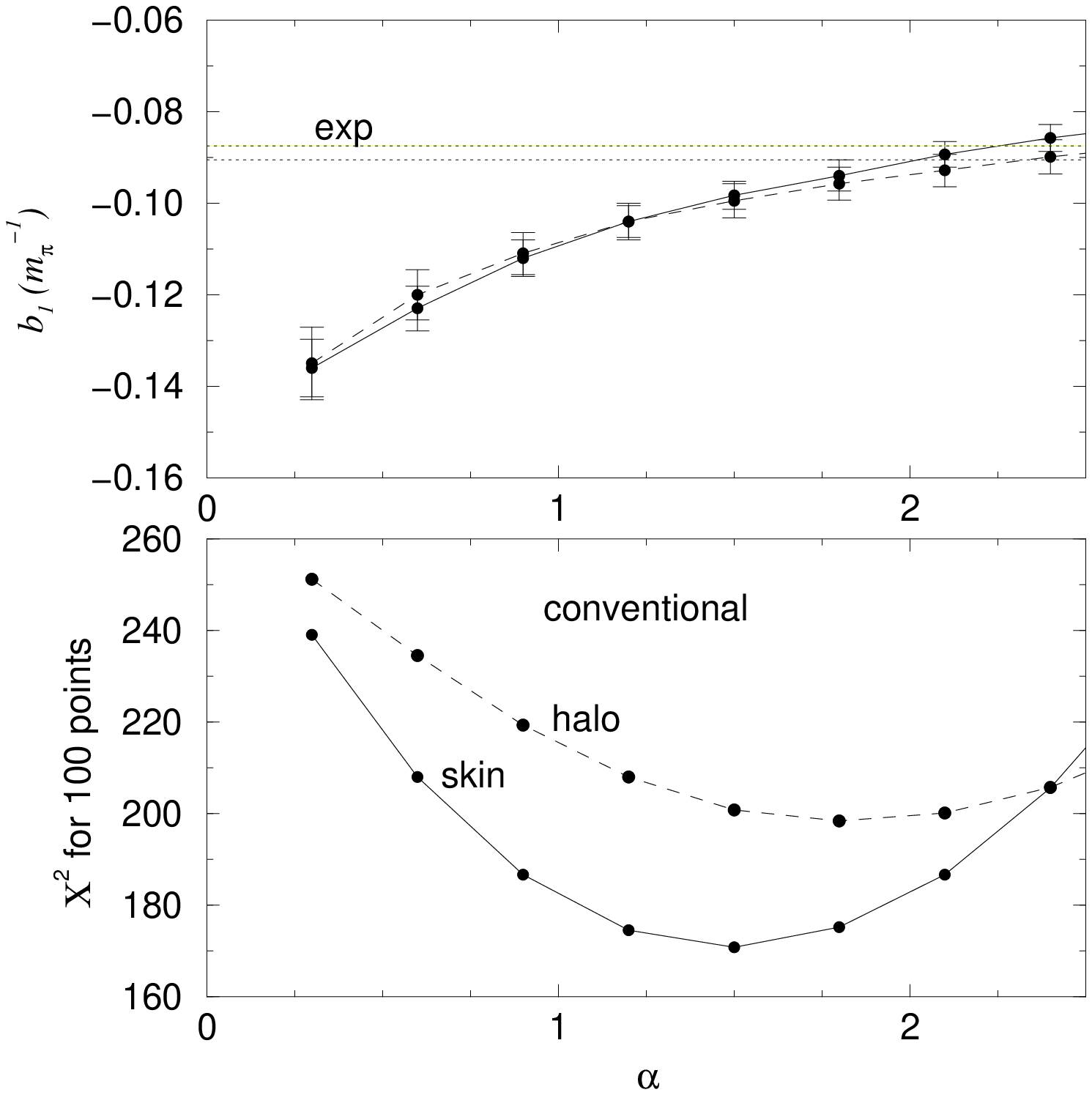, height=90mm,width=90mm}
\caption{Values of $\chi ^2$ for 100 data points (lower) and best-fit
values of $b_1$ (upper) as function of the neutron density parameter $\alpha$.
`exp' marks the experimental value for the free $\pi N$ interaction}
\label{fig:con}
\end{figure}

\begin{figure}
\epsfig{file=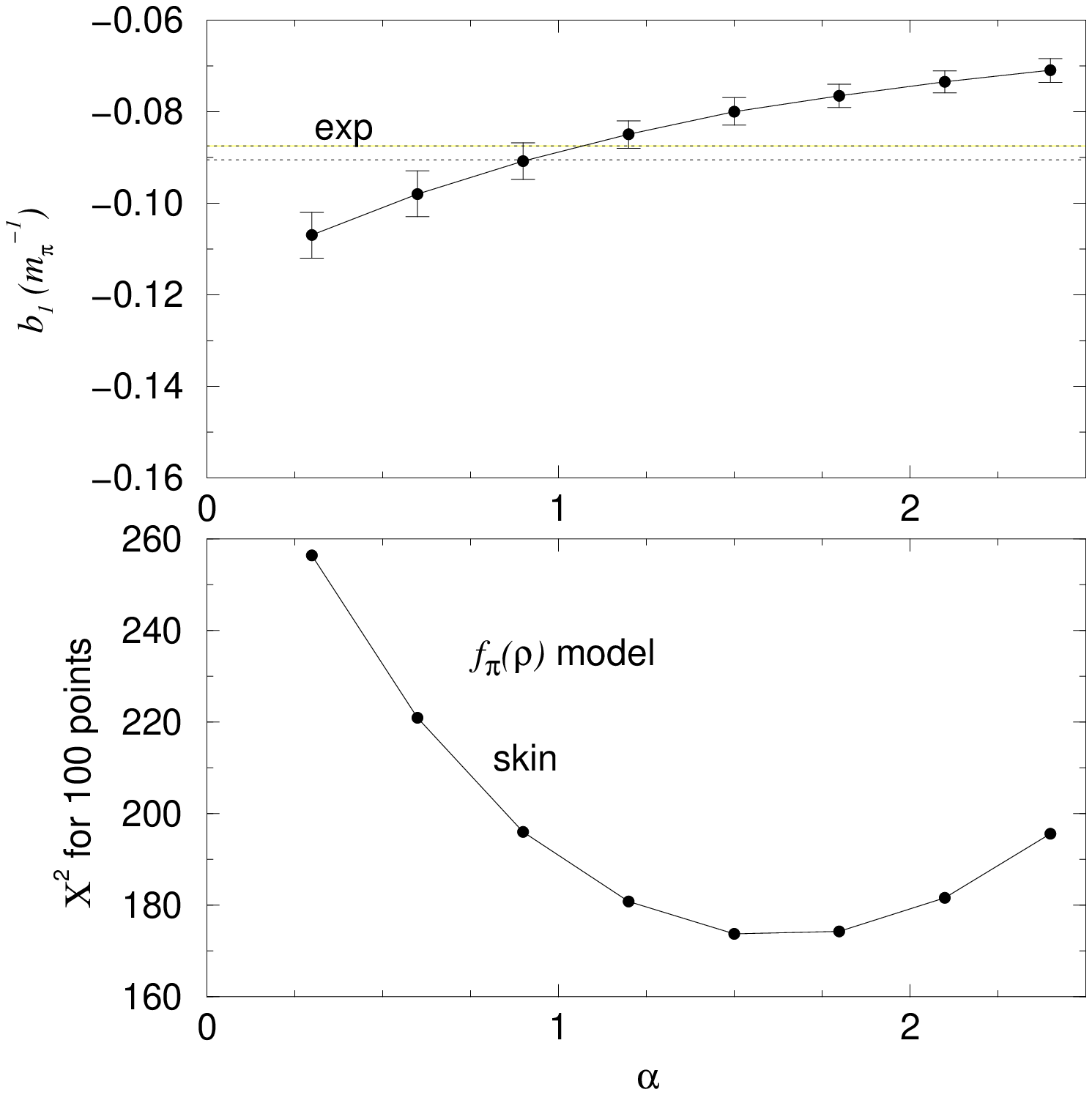, height=90mm,width=90mm}
\caption{Values of $\chi ^2$ for 100 data points (lower) and best-fit
values of $b_1$ (upper) as function of the neutron density parameter $\alpha$.
`exp' marks the experimental value for the free $\pi N$ interaction}
\label{fig:W}
\end{figure}

\begin{figure}
\epsfig{file=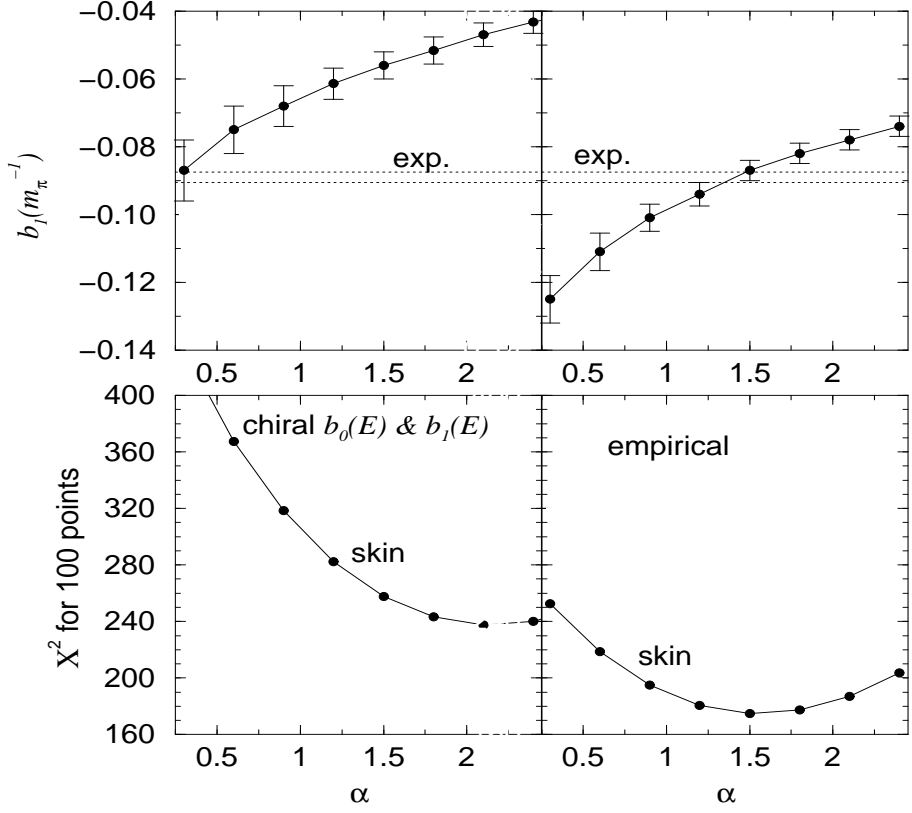, height=110mm,width=120mm}
\caption{Values of $\chi ^2$ for 100 data points (lower) and best-fit
values of $b_1$ (upper) as function of the neutron density parameter $\alpha$.
`exp' marks the experimental value for the free $\pi N$ interaction.
Left panels, for the chiral energy dependence  of $b_0$
and $b_1$; right panels, for the empirical energy dependence.}
\label{fig:E}
\end{figure}

\begin{table}
\caption{Dependence of parameter values and uncertainties on
the extent of data base and on assumptions made regarding Re$B_0$.
Deterioration in the quality of fits is measured by
 $\frac{\Delta \chi ^2}{\chi ^2 /{\rm F}}$, the increase of $\chi ^2$ in
units of $\chi ^2 $ per degree of freedom. \newline 
Values of $r_n$ are for $\alpha$=1.5}
\label{tab:stat}
\begin{tabular}{lcccc}
\\
data & `global 2'& `global 3'& light $N=Z$ & light $N=Z$ \\
 &$^{12}$C to $^{238}$U&$^{20}$Ne to $^{238}$U& + light $N>Z$&
+`deep'  \\
 & & & 1$s$ only&1$s$ only \\
points       & 120 & 100 & 22 & 20 \\  \hline
\\
$\chi ^2$ &237& 171& 54& 35 \\
 $\chi ^2 $/F  & 2.1 & 1.8 & 3.0 & 2.2 \\
$b_0 (m_\pi ^{-1})$ &~~0.000(6)&$-$0.001(7)&
$-$0.009(17)&$-$0.016(13)\\
$b_1 (m_\pi ^{-1})$ &$-$0.101(3)&$-$0.098(3)&
$-$0.095(13)&$-$0.094(7) \\
Re$B_0(m_\pi ^{-4})$&$-$0.085(30)&$-$0.082(30)&
$-$0.048(72)&$-$0.017(60)\\
Im$B_0(m_\pi ^{-4})$&~0.049(2)&~0.052(2)&
~0.049(2)&~0.051(2)\\
\\
 $\chi ^2 $ & 259 &190 & 56 & 35  \\
  $\chi ^2 $/F  & 2.3 & 2.0 & 2.9 & 2.1 \\
 $\frac{\Delta \chi ^2}{\chi ^2 /{\rm F}}$& 9.6 &10.6 & 0.7 & 0 \\
$b_0 (m_\pi ^{-1})$ &$-$0.018(1)&$-$0.019(1)&
$-$0.020(3)&$-$0.020(2)\\
$b_1 (m_\pi ^{-1})$ &$-$0.102(3)&$-$0.099(3)&
$-$0.093(12)&$-$0.094(7) \\
Re$B_0(m_\pi ^{-4})$& 0. (fixed)& 0. (fixed)& 0. (fixed)& 0. (fixed) \\
Im$B_0(m_\pi ^{-4})$&~0.048(2)&~0.051(2)&
~0.048(2)&~0.050(2)\\
\end{tabular}

`deep' refers to deeply bound 1$s$ states in $^{115,119,123}$Sn and
$^{205}$Pb.  \newline
 $b_0=-0.0001^{+0.0009}_{-0.0021} ~m_\pi ^{-1}$
and  $b_1^{}=-0.0885^{+0.0010}_{-0.0021} ~m_\pi ^{-1}$
for the free $\pi N$ interaction.

\end{table}

\end{document}